\documentclass[11pt, english]{article}
\usepackage{babel}
\pdfoutput=1
\usepackage[T1]{fontenc}
\usepackage[utf8]{inputenc}
\usepackage[a4paper]{geometry}
\usepackage{float}
\usepackage{color}
\usepackage[usenames,dvipsnames]{xcolor}
\usepackage{graphicx}
\usepackage{subfig}
\usepackage{hyperref}

\usepackage[square]{natbib}
\let\cite\citep

\hypersetup{
    bookmarks=true,         
    bookmarksnumbered=false,
    bookmarksopen=false,
    breaklinks=true,
    backref=true,
    pdftoolbar=true,        
    pdfmenubar=true,        
    pdffitwindow=false,     
    pdfstartview={FitH},    
    pdftitle={modelos paramétricos},
    pdfauthor={Fernando Antoñanzas Torres},     
    pdfcreator={AucTeX/Emacs},   
    pdfproducer={LaTeX}, 
    pdfnewwindow=true,      
    pdfborder={0 0 0},
    colorlinks=true,       
    linkcolor=BrickRed,          
    citecolor=BrickRed,        
    filecolor=black,      
    urlcolor=Blue           
}
\usepackage{longtable}
\usepackage{booktabs}
\usepackage{flafter}
\usepackage{siunitx}
\newunit{\wattpeak}{Wp}
\newunit{\watthour}{Wh}
\usepackage{mathpazo}
\usepackage{amsmath}

\usepackage{listings}

\begin{document}


  \title{Downscaling of global solar irradiation in R}

\author{
  F.~Antonanzas-Torres\footnote{\textit{Corresponding author:} \url{antonanzas.fernando@gmail.com}}\\
  Edmans Group\\
  \textit{ETSII, University of La Rioja, Logroño, Spain}
  \and
  F. J.~Martínez-de-Pisón\\
  Edmans Group\\
 \textit{ETSII, University of La Rioja, Logroño, Spain}
  \and
  J.~Antonanzas\\
  Edmans Group\\
  \textit{ETSII, University of La Rioja, Logroño, Spain}
  \and
  O.~Perpinan\\
  Electrical Engineering Department\\
  \textit{ETSIDI, Universidad Politecnica de Madrid, Spain}
}

\maketitle

  \begin{abstract}

    A methodology for downscaling solar irradiation from
    satellite-derived databases is described using \texttt{R}
    software. Different packages such as \texttt{raster},
    \texttt{parallel}, \texttt{solaR}, \texttt{gstat}, \texttt{sp} and
    \texttt{rasterVis} are considered in this study for improving
    solar resource estimation in areas with complex topography, in
    which downscaling is a very useful tool for reducing inherent
    deviations in satellite-derived irradiation databases, which lack
    of high global spatial resolution. A topographical analysis of
    horizon blocking and sky-view is developed with a digital
    elevation model to determine what fraction of hourly solar
    irradiation reaches the Earth’s surface. Eventually, kriging with
    external drift is applied for a better estimation of solar
    irradiation throughout the region analyzed. This methodology has
    been implemented as an example within the region of La Rioja in
    northern Spain, and the mean absolute error found is a striking
    25.5\% lower than with the original database.

    \textbf{Keywords: }Solar irradiation, R, \texttt{raster},
    \texttt{solaR}, digital elevation model, shade analysis,
    downscaling.
  \end{abstract}

\section{Introduction}

During the last few years the development of photovoltaic energy has
flourished in developing countries with both multi-megawatt power
plants and micro installations. However, the scarcity of long-term,
reliable solar irradiation data from pyranometers in many of these
countries makes it necessary to estimate solar irradiation from other
meteorological variables or satellite photographs
\cite{Schulz.Albert.ea2009}. In such cases, models need to be
validated via nearby pyranometer records, since they lack spatial
generalization. Thus, in some regions in which there are no
pyranometers nearby these models are ruled out as an option and
irradiation data must be obtained from satellite estimates. Although
satellite-derived irradiation databases such as NASA’s Surface
meteorology and Solar Energy
(SSE)\footnote{\url{http://maps.nrel.gov/SWERA}}, the National
Renewable Energy Laboratory
(NREL)\footnote{\url{http://www.nrel.gov/gis/solar.html}},
INPE\footnote{\url{http://www.inpe.br}},
SODA\footnote{\url{http://www.soda-is.com/eng/index.html}} and the
Climate Monitoring Satellite Application Facility (CM
SAF)\footnote{\url{http://www.cmsaf.eu}} provide wide spatial
coverage, only NASA and some CM SAF climate data sets give global
coverage, albeit at a reduced spatial resolution
(Table~\ref{tab:databases}).  

\begin{small}
\begin{longtable}{p{0.12\textwidth}p{0.16\textwidth}p{0.16\textwidth}p{0.10\textwidth}p{0.12\textwidth}p{0.12\textwidth}}
\toprule
\ Database&Product&Spatial coverage&Spatial resolution&Temporal coverage&Temporal resolution \\
\midrule
\endhead
CM~SAF&SIS Climate Data Set (GHI)&Global&0.25x0.25$^\circ$&1982-2009&Daily means\\
\midrule
CM~SAF&SIS Climate Data Set (GHI)&70S-70N, 70W-70E&0.03x0.03$^\circ$&1983-2005&Hourly means\\
\midrule
CM~SAF&SID Climate Data Set (BHI)&70S-70N, 70W-70E&0.03x0.03$^\circ$&1983-2005&Hourly means\\
\midrule
SODA& Helioclim 3 v2 and v3 (GHI)&66S-66N,66W-66E&~5km&2005&15 minutes\\
\midrule
SODA& Helioclim 3 v2 and v3 (GHI)&66S-66N,66W-66E&~5km&2005&15 minutes\\
\midrule
NREL& GHI Moderate resolution&Central and South America, Africa, India, East Asia&40x40km&1985-1991&Monthly means of daily GHI\\
\midrule
NASA& SSE&Global&1x1$^\circ$&1983-2005&Daily means\\
\bottomrule
\caption{Summary of solar irradiation databases}
\label{tab:databases}
\end{longtable}
\end{small}

The spatial resolutions of satellite
estimates are generally in the range of kilometers: they tend to
average solar irradiation and omit the impact of topography within
each cell. As a result, intra-cell variations can be very significant
in areas with local micro-climatic characteristics and in areas with
complex topography (which are often one and the same). In this case,
the irradiation data might not be accurate enough to enable a
photovoltaic installation to be designed. \cite{Perez.Seals.ea1994}
analyze the spatial behavior of solar irradiation and conclude that
the break-even distance from satellite estimates to pyranometers is in
the order of 7 km and that variations are hard to estimate for
distances greater than 40
km. \cite{Antonanzas-Torres.Canizares.ea2013} reject ordinary kriging
as a spatial interpolation method for solar irradiation in Spain with
stations more than 50 km apart in mountainous regions, as a result of
the high spatial variability in such areas. The NASA-SSE and CM SAF
SIS Climate Data Sets (GHI) provide global coverage with resolutions
of 1x1$^\circ$ and 0.25x0.25$^\circ$ (Table~\ref{tab:databases}),
which in most latitudes implies a grosser resolution than the
previously mentioned 40-50 km.

One of the alternatives for obtaining higher spatial resolution of
solar irradiation is the downscaling of satellite
estimates. Irradiation downscaling can be based on interpolation
kriging techniques when pyranometer records are available, with the
implementation of continuous irradiation-related variables such as
elevation, sky-view-factor and other meteorological variables as
external drifts \cite{Alsamamra.Ruiz-Arias.ea2009,
  Batlles.Bosch.ea2008}. Downscaling is generally based on digital
elevation models (DEM) with satellite-derived irradiation data to
account for the effect of complex topography. It has previously been
applied in mountainous areas such as the Mont Blanc Massif (France)
\cite{Corripio2003} and Sierra Nevada (Spain)
\cite{Bosch.Batlles.ea2010,Ruiz-Arias.Cebecauer.ea2010} with image
resolutions of 3.5x3.5 km. However, the NASA-SSE and CM SAF SIS
Climate Data Sets are based on much lower resolutions and are the only
irradiation datasets in numerous countries where there has been recent
interest in solar energy. In this paper, a downscaling methodology of
global solar irradiation is explained by means of \texttt{R} software
and studied in the region of La Rioja (a very mountainous region in
northern Spain). Data from the CM SAF with 0.03x0.03$^\circ$
resolution is considered and then downscaled to a higher resolution
(200x200 m). In a second step, \emph{kriging with external drift},
also referred to as \emph{universal kriging}, is applied to
interpolate data from 6 on-ground pyranometers in the region, and this
downscaled CM SAF data is considered as an explanatory
variable. Finally, a downscaled map of annual global solar radiation
throughout this region is obtained.

\section{Data}

\begin{figure}[H]
\centering
\includegraphics[width=0.9\textwidth]{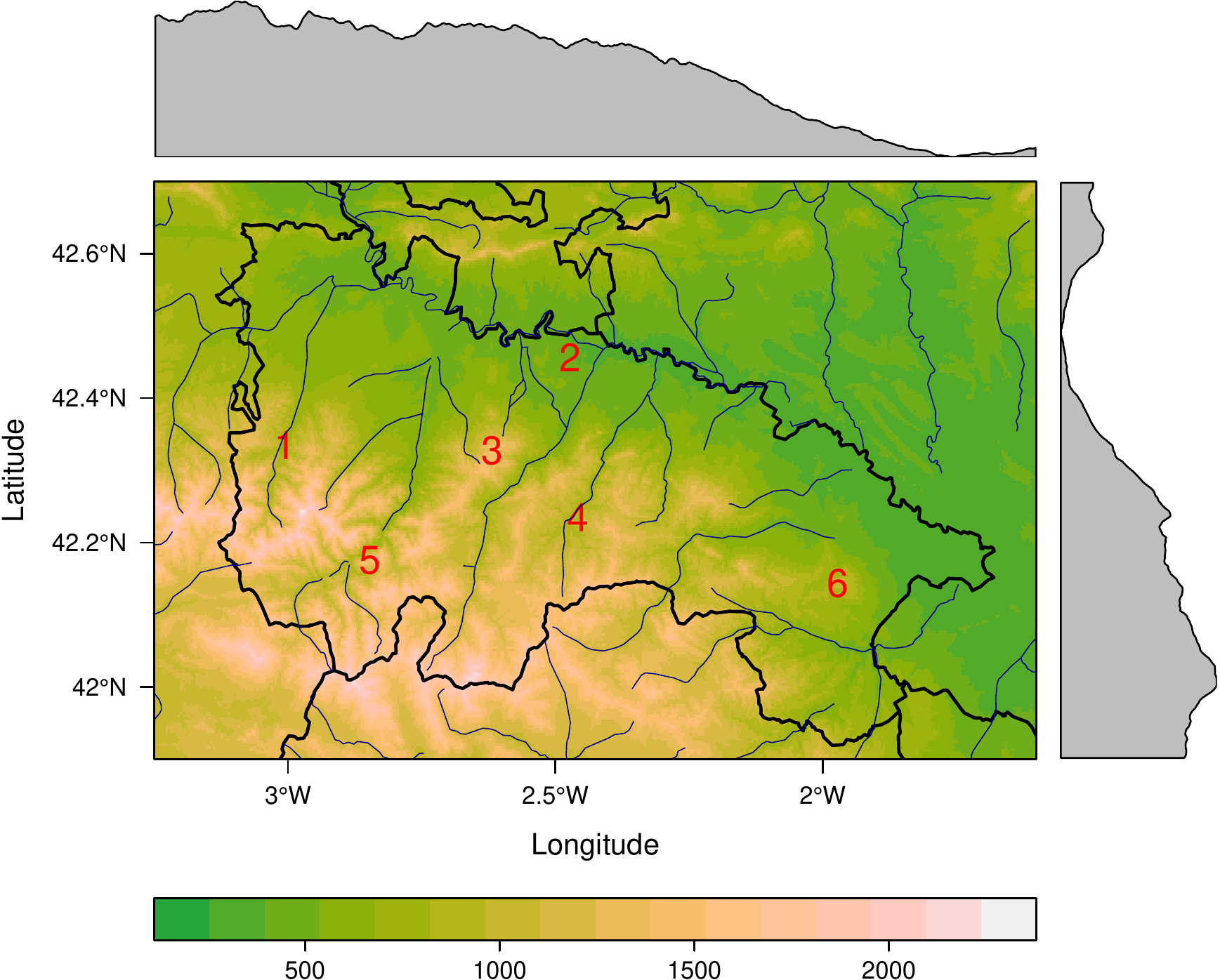}
\caption{Region analyzed and meteorological stations considered}
\label{fig:mapstations}
\end{figure}

The CM SAF was funded in 1992 as a joint venture of several European
meteorological institutes, with the collaboration of the European
Organization for the Exploitation of Meteorological Satellites
(EUMETSAT) to retrieve, archive and distribute climate data to be used
for climate monitoring and climate analysis
\cite{Posselt.Mueller.ea2012a}. Two categories are provided:
operational products and climate data. Operational products are built
on data validated with on-ground stations and provided in
near-to-present time and climate data are long-term series for
evaluating inter-annual variability. This study is built on hourly
surface incoming solar radiation and direct irradiation climate data,
denoted as SIS and SID by CM SAF respectively, for the year
2005. These climate data are derived from Meteosat first generation
satellites (Meteosat 2 to 7, 1982-2005) and validated using on-ground
records from the Baseline Surface Radiation Network (BSRN) as a
reference. The target accuracy of SIS and SID in hourly means is 15
$W/m^{2}$ \cite{Posselt.Muller.ea2011}, providing a maximum spatial
resolution of 0.03x0.03$^\circ$. In the study, SIS and SID data are
selected with spatial resolution of 0.03x0.03$^\circ$. Data is freely
accessible via FTP through the CM SAF website.  Hourly GHI records
from SOS
Rioja\footnote{\url{http://www.larioja.org/npRioja/default/defaultpage.jsp?idtab=442821}},
taken from 6 meteorological stations (shown in
Figure~\ref{fig:mapstations} and Table~\ref{tab:stations}) in 2005
serve as complementary measurements for downscaling within the region
studied.  These stations have First Class pyranometers (according to
ISO 9060) with uncertainty levels of 5\% in daily totals. These data
are filtered from spurious, assuming when relevant the average between
the previous and following hourly measurements.  The digital elevation
model (DEM) is also freely obtained from product MDT-200 by the
\copyright Spanish Institute of
Geography\footnote{\url{http://www.ign.es}} with a spatial resolution
of 200x200 m.

\begin{small}
\begin{longtable}{p{0.01\textwidth}p{0.12\textwidth}p{0.06\textwidth}p{0.05\textwidth}p{0.06\textwidth}p{0.05\textwidth}p{0.05\textwidth}}
\toprule
\# & Name & Net. & Lat.(º) & Long.(º) & Alt. & $GHI_{a}$\\
\midrule
\endhead
1 & Ezcaray & SOS & 42.33 & -3.00 & 1000 & 1479\\
\midrule
2 & Logroño & SOS & 42.45 & -2.74 & 408 & 1504\\
\midrule 
3 & Moncalvillo & SOS & 42.32 & -2.61 & 1495 & 1329 \\
\midrule
4 & San Roman & SOS & 42.23 & -2.45 & 1094 & 1504 \\
\midrule
5 & Ventrosa & SOS & 42.17 & -2.84 & 1565 & 1277\\
\midrule
6 & Yerga & SOS& 42.14 & -1.97 & 1235 &1448\\
\bottomrule
\caption{Summary of the meteorological
stations selected.}
\label{tab:stations}
\end{longtable}
\end{small}

\section{Method}
\label{sec:method}

\begin{figure}
\centering
\includegraphics[width=\textwidth]{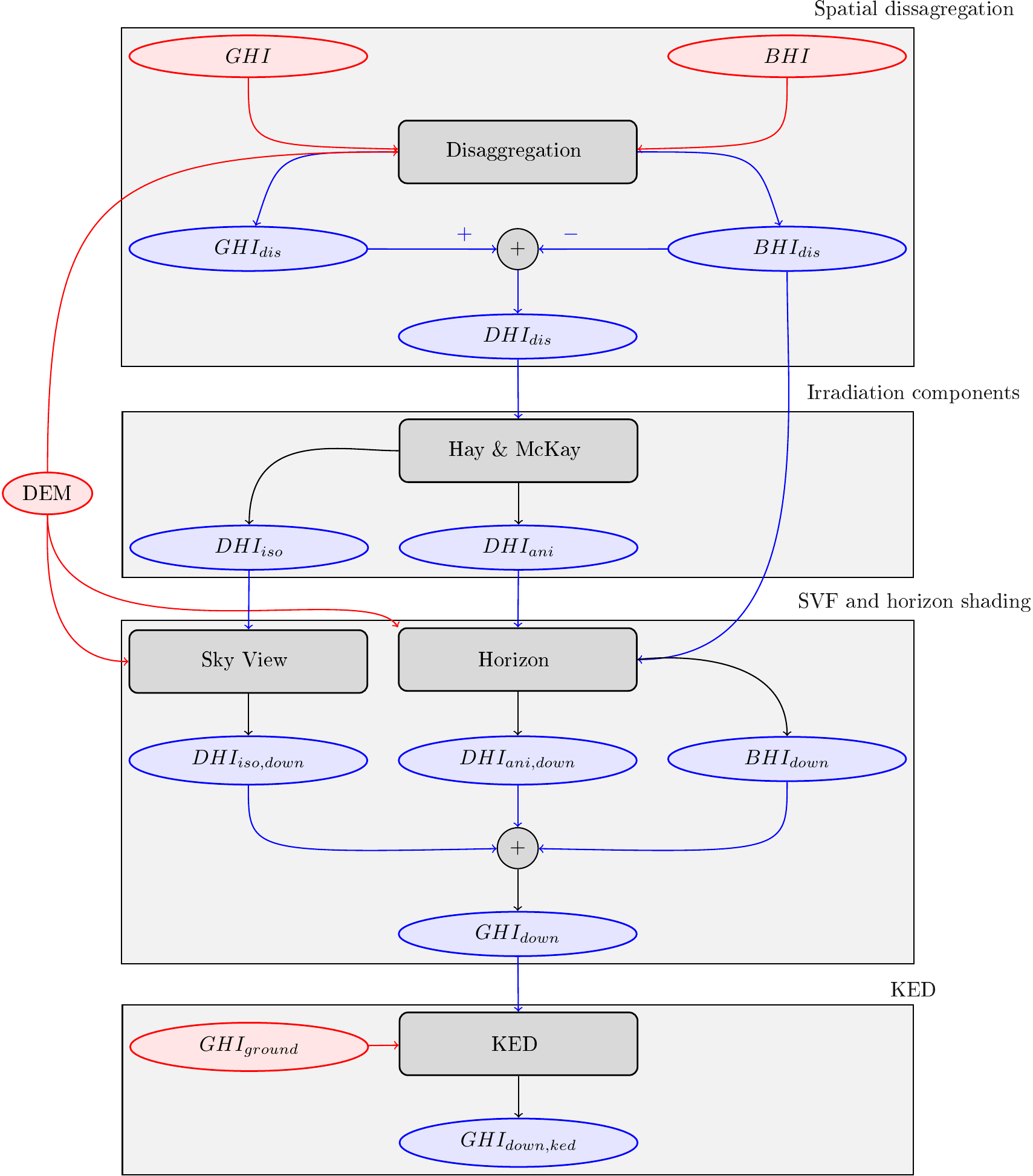}
\caption{Methodology of downscaling: this figure uses red ellipses
and lines for data sources, blue ellipses and lines for derived
rasters (results), and black rectangles and lines for operations.
}
\label{fig:method}
\end{figure}

This section describes the methodology proposed. Figure~\ref{fig:method} displays the method diagram using red ellipses and lines for data sources, blue ellipses and lines for derived rasters (results), and black rectangles and lines for operations.

\subsection{Irradiation decomposition}
\label{sec:irradDecomp}

Initially, diffuse horizontal irradiation (\emph{DHI}) is obtained
from the difference between global horizontal irradiation (\emph{GHI})
and beam horizontal irradiation (\emph{BHI}) rasters, previously
obtained from CM SAF. \emph{DHI} and \emph{BHI} are firstly
disaggregated from the original gross resolution (0.03x0.03$^\circ$)
into the DEM resolution (200x200 m), leading to similar values
remaining in disaggregated pixels to the original gross resolution
pixel. In a second step, \emph{DHI} can be divided in two components:
isotropic diffuse irradiation ($DHI_{iso}$), and anisotropic diffuse
irradiation ($DHI_{ani}$) as per the model by Hay \& Mckay
\cite{Hay.Mckay1985} (Equation~\ref{eq:DHI}). This model is based on
the anisotropy index ($k_1$), defined as the ratio of the beam
irradiance ($B(0)$) to the extra-terrestrial irradiance ($B_{0}(0)$),
as shown in Equation~\ref{eq:k1}. High $k_1$ values are typical in
clear sky atmospheres, while low $k_1$ values are frequent in overcast
atmospheres and those with a high aerosol density.

\begin{equation}
\label{eq:DHI}
DHI=DHI_{iso}+DHI_{ani}
\end{equation}
\begin{equation}
\label{eq:k1}
k_{1}=\frac{B(0)}{B_{0}(0)} 
\end{equation}

The $DHI_{iso}$ accounts for the incoming diffuse irradiation portion from an isotropic sky, and is more significant on very cloudy days (Equation~\ref{eq:DHIisos}).

\begin{equation}
\label{eq:DHIisos}
DHI_{iso}=DHI\cdot{}(1-k_{1})
\end{equation}
 
$DHI_{ani}$, also denoted as circumsolar diffuse irradiation, considers the incoming portion from the circumsolar disk and can be analyzed as beam irradiation \cite{Perpinan-Lamigueiro2013} (Equation ~\ref{eq:DHIani}).

\begin{equation}
\label{eq:DHIani}
DHI_{ani}=DHI\cdot{}k_{1}
\end{equation}

\subsection{Sky view factor and horizon blocking}
\label{sec:sky-view-factor}

Topographical analysis is performed accounting for the visible sky
sphere (sky view) and horizon blocking. The $DHI_{iso}$ is directly
dependent on the sky-view factor (SVF), which computes the proportion
of visible sky related to a flat horizon. The sky-view factor is
considered in earlier irradiation assessments
\cite{Ruiz-Arias.Cebecauer.ea2010, Corripio2003}. It is calculated in
each DEM pixel by considering 72 vectors (separated by 5$^\circ$ each)
and evaluating the maximum horizon angle ($\theta_{hor}$) over 20 km
in each vector (Equation~\ref{eq:SVF}). The $\theta_{hor}$ stands for
the maximum angle between the altitude of a location and the elevation
of the group of points along each vector, related to a horizontal
plane on the location. Locations without horizon blocking have SVFs
close to 1, which means a whole visible semi-sphere of sky.

\begin{equation}
\label{eq:SVF}
SVF=1-\int_0^{2\pi}sin^{2} \theta_{hor} d\theta
\end{equation}

Eventually, the downscaled $DHI_{iso}$ ($DHI_{iso,down}$) is
computed with Equation~\ref{eq:DHIiso}.

\begin{equation}
\label{eq:DHIiso}
DHI_{iso,down}=DHI_{iso}\cdot{}SVF
\end{equation}

Horizon blocking is analyzed by evaluating the solar geometry in 15 minute samples, particularly the solar elevation ($\gamma_{s}$) and the solar azimuth ($\psi_{s}$). Secondly, the mean hourly $\gamma_{s}$ and $\psi_{s}$ (from those 15 minute rasters) are calculated and then disaggregated as explained above for \emph{DHI} and \emph{BHI} rasters. The decision to solve the solar geometry with low resolution rasters enables computation time to be reduced significantly without penalizing the results. The $\theta_{hor}$ corresponding to each $\psi_{s}$ is compared with the $\theta_{zs}$. As a consequence, if the $\theta_{zs}$ is greater than the $\theta_{hor}$, then there is horizon blocking on the surface analyzed and therefore, \emph{BHI} and $DHI_{ani}$ are blocked. Finally, the sum of $DHI_{ani,down}$, $DHI_{iso,down}$ and $BHI_{iso,down}$ constitutes the downscaled global horizontal irradiation $GHI_{down}$.

\subsection{Post-processing: kriging with external drift}
\label{sec:meth}

The fact that this downscaling accounts for the irradiation loss due
to horizon blocking and the sky-view factor leads us to introduce a
trend in estimates (lowering them) compared to the original data
(gross resolution data). However, satellite-derived irradiation data
implicitly considers shade, as a consequence of the lower albedo
recorded in these zones, although it is later averaged over the
pixel. $GHI_{down}$ can be considered as a useful bias of the behavior
of solar irradiation within the region studied. \emph{Universal
  kriging} or \emph{kriging with external drift} (KED) includes
information from exhaustively-sampled explanatory variables in the
interpolation. As a result, $GHI_{down}$ is considered as the
explanatory variable for interpolating measured irradiation data from
on-ground calibrated pyranometers, which is denoted as
\emph{post-processing}. $GHI_{down}$ is correlated with the DEM as a
consequence of the major influence of horizon blocking with
topography, estimations can be derived by separating the deterministic
($\hat{m}(\mathbf{s}_\theta)$) and stochastic components
($\hat{\epsilon}(\mathbf{s}_\theta)$ (Equations~\ref{eq:externalDrift}
and ~\ref{eq:externalDrift_sum}).

\begin{equation}
\label{eq:externalDrift}
\hat{z}(\mathbf{\mathbf{s}}_\theta) = \hat{m}(\mathbf{s}_\theta) + \hat{\epsilon}(\mathbf{s}_\theta)
\end{equation}
\begin{equation}
\label{eq:externalDrift_sum}
\hat{z}(\mathbf{s}_\theta) = \sum_{k=0}^p \hat{\beta}_k q_k(\mathbf{s}_\theta) + 
\sum_{i=1}^n \lambda_i \epsilon(\mathbf{s}_i)
\end{equation}

where $\hat{\beta}_k$ are the estimated coefficients of the
deterministic model, $q_k(\mathbf{s}_\theta)$ are the auxiliary
predictors obtained from the fitted values of the explanatory variable
at the new location, $\lambda_i$ are the kriging weights determined by
the spatial dependence structure of the residual, and
$\epsilon(\mathbf{s}_i)$ are the residual at location $\mathbf{s}_i$
\cite{Antonanzas-Torres.Canizares.ea2013}.

The semivariogram is a function defined as Equation~\ref{eq:variogram}
based on a constant variance of $\epsilon$ and also on the assumption
that spatial correlation of $\epsilon$ depends on the distance amongst
instances ($\mathbf{h}$) rather than on their position
\cite{Pebesma2004}.

\begin{equation}
\label{eq:variogram}
\gamma(\mathbf{h}) = \frac{1}{2} \textrm{E}(\epsilon(\mathbf{s}) -
\epsilon(\mathbf{s} + \mathbf{h}))^2
\end{equation}

Given that the above sample variogram only collates estimates from
observed points, a fitting model of this variogram is generally
considered to extrapolate the spatial behavior of observed points to
the area studied. In the literature different variogram functions are
commonly defined such as the exponential, Gaussian or spherical
models. Along these lines, different parameters such as the sill,
range and nugget of the model must be adjusted to best fit the sample
variogram \cite{Hengl2009}. The nugget effect, generally associated
with intrinsic micro-variability and measurement error, models the
discontinuity of the variogram at the source. It must be highlighted
that when the nugget effect is recorded, kriging differs from a
regular interpolation and as a result estimates are different from
measured values. The variogram model of solar horizontal irradiation
is evaluated in Spain, and the conclusion reached is that a pure
nugget fitting behaves best, which implies no spatial auto-correlation
on residuals \cite{Antonanzas-Torres.Canizares.ea2013}.

\section{Implementation}
\label{sec-1}

\begin{figure}[H]
  \centering
  \includegraphics[width=0.9\textwidth]{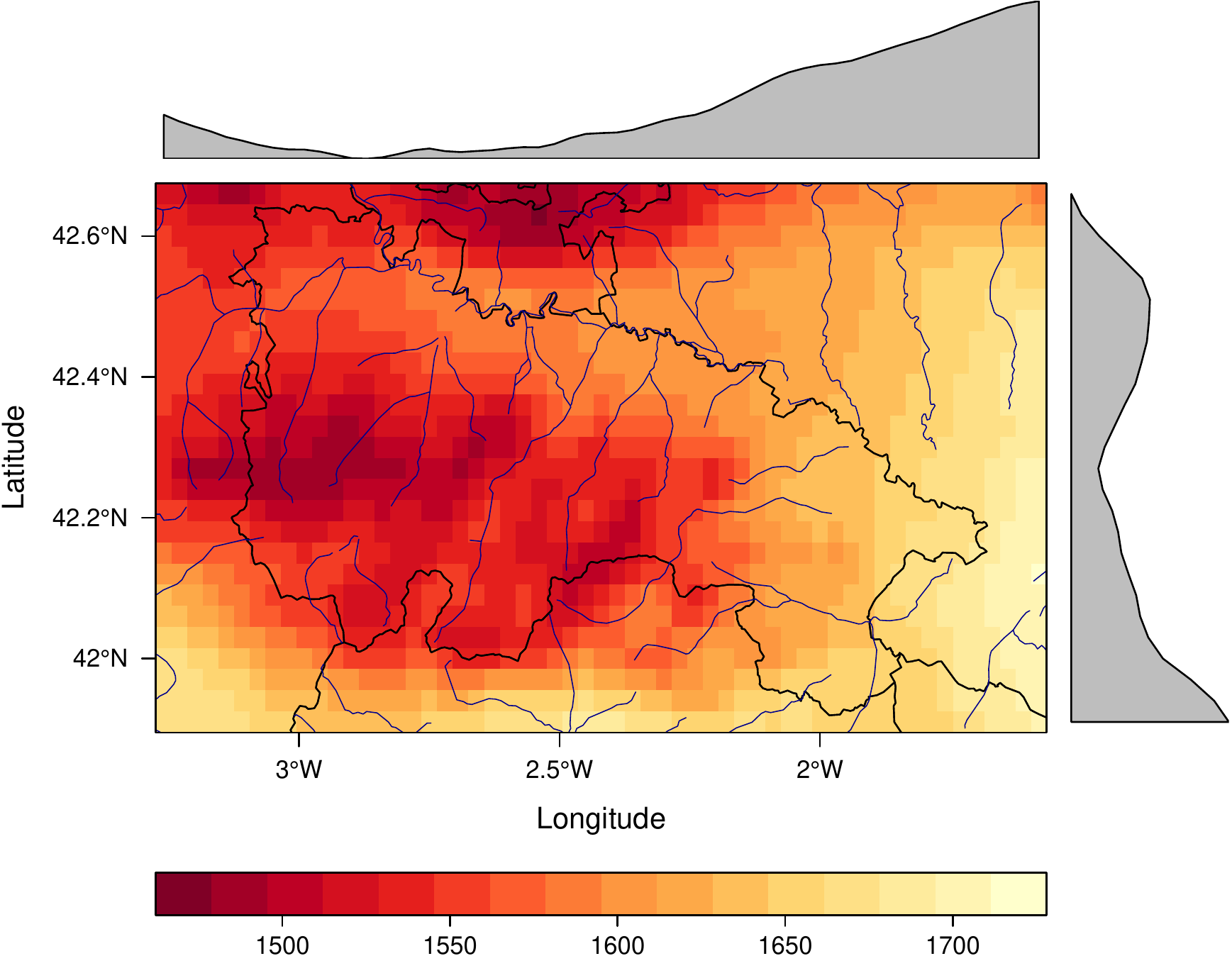}
  \caption{Annual GHI of 2005 from CM~SAF estimates (0.03x0.03$^\circ$) in La Rioja}
  \label{fig:cmsaf}
\end{figure}

The method proposed is applied in the region of La Rioja (northern
Spain). Figure \ref{fig:cmsaf} shows the corresponding annual global
horizontal irradiation from CM SAF with resolution 0.03x0.03$^\circ$.

\subsection{Packages}
\label{sec-1-1}

The downscaling described in this paper has been implemented using the
free software environment \texttt{R}
\cite{proglangRDevelopmentCoreTeam2013} and various contributed
packages:

\begin{itemize}
\item \texttt{raster} \cite{Hijmans.Etten2013} for spatial data manipulation
and analysis.
\item \texttt{solaR} \cite{Perpinan-Lamigueiro2012} for solar
geometry.
\item \texttt{gstat} \cite{Pebesma.Graeler2013} and \texttt{sp}
\cite{Pebesma.Bivand.ea2013} for geostatistical analysis.
\item \texttt{parallel} for multi-core parallelization.
\item \texttt{rasterVis} \cite{Perpinan-Lamiguiero.Hijmans2013} for spatial data
visualization methods.
\end{itemize}

\lstset{language=R,numbers=none}
\begin{lstlisting} 
R> library(sp)
R> library(raster)
R> rasterOptions(todisk=FALSE)
R> rasterOptions(chunksize = 1e+06, maxmemory = 1e+07)
R> library(maptools)
R> library(gstat)
R> library(lattice)
R> library(latticeExtra)
R> library(rasterVis)
R> library(solaR)
R> library(parallel)
\end{lstlisting}

\subsection{Data}
\label{sec-1-2}

Satellite data can be freely downloaded after registration from CM
SAF\footnote{\url{www.cmsaf.eu}} by going to the data access area,
selecting \emph{web user interface} and \emph{climate data sets} and
then choosing the hourly climate data sets named \emph{SIS} (Global
Horizontal Irradiation)) and \emph{SID} (Beam Horizontal Irradiation)
for 2005. Both rasters are projected to the UTM projection for
compatibility with the DEM.

\lstset{language=R,numbers=none}
\begin{lstlisting} 
R> projUTM  <-  CRS('+proj=utm +zone=30')
R> projLonLat <- CRS(' +proj=longlat +ellps=WGS84')

R> listFich <- dir(pattern='SIShm2005')
R> stackSIS <- stack(listFich)
R> stackSIS <- projectRaster(stackSIS,crs=projUTM)

R> listFich <- dir(pattern='SIDhm2005')
R> stackSID <- stack(listFich)
R> stackSID <- projectRaster(stackSID, crs=projUTM)
\end{lstlisting}

We compute the annual global irradiation, which will be used as a reference for subsequent steps.

\lstset{language=R,numbers=none}
\begin{lstlisting} 
R> SISa2005 <- calc(stackSIS, sum, na.rm=TRUE)
\end{lstlisting}

The Spanish Digital Elevation Model can be obtained after registration
from the \copyright Spanish Institute of
Geography\footnote{\url{http://www.ign.es}} by going to the\emph{free
  download of digital geographic information for non-commercial use}
area, and then cropping to the region analyzed (La Rioja). As stated
above, this DEM uses the UTM projection.

\lstset{language=R,numbers=none}
\begin{lstlisting} 
R> elevSpain <- raster('elevSpain.grd')
R> elev <- crop(elevSpain, extent(479600, 616200, 4639600, 4728400))
R> names(elev)<-'elev'
\end{lstlisting}

\subsection{Sun geometry}
\label{sec-1-3}

The first step is to compute the sun angles (height and azimuth) and
the extraterrestrial solar irradiation for each cell of the CM SAF
rasters. The function \texttt{calcSol} from the \texttt{solaR} package
calculates the daily and intradaily sun geometry. By means of this
function and \texttt{overlay} from the \texttt{raster} package, three
multilayer \texttt{raster} objects are generated with the sun geometry
needed for the next steps. For the sake of brevity we show only the
procedure for extraterrestrial solar irradiation. The sun geometry is
calculated with the resolution of CM SAF.  First, it is defined a
function to extract the hour for aggregation, choose the annual
irradiation raster as reference, and define a raster with longitude
and latitude coordinates.

\lstset{language=R,numbers=none}
\begin{lstlisting} 
R> hour <- function(tt)as.POSIXct(trunc(tt, 'hours'))

R> r <- SISa2005

R> latlon <- stack(init(r, v='y'), init(r, v='x'))
R> names(latlon) <- c('lat', 'lon')
\end{lstlisting}

The extraterrestrial irradiation is calculated with 5-min
samples. Each point is a column of the \texttt{data frame locs}. Its
columns are traversed with \texttt{lapply}, so for each point of the
\texttt{raster} object a time series of extraterrestrial solar
irradiation is computed. The result, \texttt{B05min}, is a
\texttt{RasterBrick} object with a layer for each element of the time
index \texttt{BTi}, which is aggregated to an hourly raster with
\texttt{zApply} and transformed to the UTM projection.

\lstset{language=R,numbers=none}
\begin{lstlisting} 
R> BTi <- seq(as.POSIXct('2005-01-01 00:00:00'),
+            as.POSIXct('2005-12-31 23:55:00'), by='5 min')

R> B05min <- overlay(latlon, fun=function(lat, lon){
+     locs <- as.data.frame(rbind(lat, lon))
+     b <- lapply(locs, function(p){
+ 
+         hh <- local2Solar(BTi, p[2])
+         sol <- calcSol(p[1], BTi=hh)
+         Bo0 <- as.data.frameI(sol)$Bo0
+         Bo0 })
+     res <- do.call(rbind, b)})

R> B05min <- setZ(B05min, BTi)
R> names(B05min) <- as.character(BTi)

R> B0h <- zApply(B05min, by=hour, fun=mean)
R> projectRaster(B0h,crs=projUTM)
\end{lstlisting}

\subsection{Irradiation components}
\label{sec-1-4}

The CM SAF rasters must be transformed to the higher resolution of the
DEM (UTM 200x200 m). Because of the differences in pixel geometry
between DEM (square) and irradiation rasters (rectangle) the process
is performed in two steps.

The first step increases the spatial resolution of the irradiation
rasters to a similar and also larger pixel size than the DEM with
disaggregated data, where \texttt{sf} is the scale factor. The second
step post-processes the previous step by means of a bilinear
interpolation which resamples the raster layer and achieves the same
DEM resolution (\texttt{resample}). This two-step disaggregation
prevents the loss of the original values of the gross resolution
raster that would be directly interpolated with the one-step
disaggregation.

\lstset{language=R,numbers=none}
\begin{lstlisting} 
R> sf <- res(stackSID)/res(elev)

R> SIDd <- disaggregate(stackSID, sf)
R> SIDdr <- resample(SIDd, elev)

R> SISd <- disaggregate(stackSIS, sf)
R> SISdr <- resample(SISd, elev)
\end{lstlisting}

On the other hand, the diffuse irradiation is obtained from the global
and beam irradiation rasters. The two components of the diffuse
irradiation, isotropic and anisotropic, can be separated with the
anisotropy index, computed as the ratio between beam and
extraterrestrial irradiation.

\lstset{language=R,numbers=none}
\begin{lstlisting} 
R> Difdr <- SISdr - SIDdr

R> B0hd <- disaggregate(B0h, sf)
R> B0hdr <- resample(B0hd, elev)

R> k1 <- SIDdr/B0hdr

R> Difiso <- (1-k1) * Difdr
R> Difani <- k1 * Difdr
\end{lstlisting}

\subsection{Sky view factor and horizon blocking}
\label{sec-1-5}

\subsubsection{Horizon angle}
\label{sec-1-5-1}

The maximum horizon angle required for the horizon blocking analysis
and to derive the SVF is obtained with the next code.  The
\texttt{alpha} vector is visited with \texttt{mclapply} (using
parallel computing). For each direction angle (elements of this
vector) the maximum horizon angle is calculated for a set of points
across that direction from each of the locations defined in
\texttt{xyelev} (derived from the DEM raster and transformed in the
matrix \texttt{locs} visited by rows).

\lstset{language=R,numbers=none}
\begin{lstlisting} 
R> xyelev <- stack(init(elev, v='x'),
+                 init(elev, v='y'),
+                 elev)
R> names(xyelev) <- c('x', 'y','elev')


R> inc <- pi/36
R> alfa <- seq(-0.5*pi,(1.5*pi-inc), inc)

R> locs <- as.matrix(xyelev)
\end{lstlisting}

Separations between the source locations and points along each
direction are defined by \texttt{resD}, the maximum resolution of the
DEM, \texttt{d}, maximum distance to visit, and consequently in the
vector \texttt{seps}.

\lstset{language=R,numbers=none}
\begin{lstlisting} 
R> resD <- max(res(elev))

R> d <- 20000
R> seps <- seq(resD, d, by=resD)
\end{lstlisting}

The elevation (\texttt{z1}) of each point in \texttt{xyelev} is
converted into the horizon angle: the largest of these angles is the
horizon angle for that direction. The result of each \texttt{apply}
step is a matrix, which is used to fill in a RasterLayer
(\texttt{r}). The result of \texttt{mclapply} is a list, \texttt{hor},
of \texttt{RasterLayer} which can be converted into a
\texttt{RasterStack} with \texttt{stack}. Each layer of this
\texttt{RasterStack} corresponds to a different direction.

\lstset{language=R,numbers=none}
\begin{lstlisting} 
R> hor <- mclapply(alfa, function(ang){
+     h <- apply(locs, 1, function(p){
+         x1 <- p[1]+cos(ang)*seps
+         y1 <- p[2]+sin(ang)*seps
+         p1 <- cbind(x1,y1)
+         z1 <- elevSpain[cellFromXY(elevSpain,p1)]
+         hor <- r2d(atan2(z1-p[3], seps))
+         maxHor <- max(hor[which.max(hor)], 0)
+     })
+     r <- raster(elev)
+     r[] <- matrix(h, nrow=nrow(r), byrow=TRUE)
+     r}, mc.cores=8)
R> horizon <- stack(hor)
\end{lstlisting}

This operation is very time-consuming as it is necessary to work with
high resolution files. Computation time can be decreased by increasing
the sampling space (200 m) or the sectoral angle (5 $^\circ$) or by
reducing the maximum distance (20 km).

\subsubsection{Horizon blocking}
\label{sec-1-5-2}

Horizon blocking is analyzed by evaluating the solar geometry in 15
minute samples, particularly the solar elevation and azimuth angles
from the original irradiation raster. Secondly, the hourly averages
are calculated, disaggregated and post-processed as explained above
for the irradiation rasters. The decision to solve the solar geometry
with low resolution rasters enables a significant reduction to be
obtained in computation time without penalizing the results.

First, the azimuth raster is cut into different classes according to
the alpha vector (directions). The values of the \texttt{horizon}
raster corresponding to each angle class are extracted using
\texttt{stackSelect}.

\lstset{language=R,numbers=none}
\begin{lstlisting} 
R> idxAngle <- cut(AzShr, breaks=r2d(alfa))
R> AngAlt <- stackSelect(horizon, idxAngle)
\end{lstlisting}

The number of layers of \texttt{AngAlt} is the same as
\texttt{idxAngle} and can therefore be used for comparison with the
solar height angle, \texttt{AlShr}. If \texttt{AngAlt} is greater,
there is horizon blocking (\texttt{dilogical=0}).

\lstset{language=R,numbers=none}
\begin{lstlisting} 
R> dilogical <- ((AngAlt-AlShr) < 0)
\end{lstlisting}

With this binary raster, beam irradiation and diffuse anisotropic
irradiation can be corrected with horizon blocking.
\lstset{language=R,numbers=none}
\begin{lstlisting} 
R> Dirh <- SIDdr * dilogical
R> Difani <- Difani * dilogical
\end{lstlisting}

\subsubsection{Sky view factor}
\label{sec-1-5-3}

The sky-view factor can be easily computed from the \texttt{horizon}
object with the equation proposed above. This factor corrects the
isotropic component of the diffuse irradiation.

\lstset{language=R,numbers=none}
\begin{lstlisting} 
R> SVFRuizArias <- calc(horizon, function(x) sin(d2r(x))^2)
R> SVF <- 1 - mean(SVFRuizArias)

R> Difiso <- Difiso * SVF
\end{lstlisting}

Finally, the global irradiation is the sum of the three corrected
components, beam and anisotropic diffuse irradiation including horizon
blocking, and isotropic diffuse irradiation with the sky view factor.

\lstset{language=R,numbers=none}
\begin{lstlisting} 
R> GHIh <- Difanis + Difiso + Dirh
R> GHI2005a <- calc(GHIh, fun=sum)
\end{lstlisting}

\subsection{Kriging with external drift}
\label{sec-1-6}

The downscaled irradiation rasters can be improved by using kriging
with external drift. Irradiation data from on-ground meteorological
stations is interpolated with the downscaled irradiation raster as the
explanatory variable. To define the variogram here we use the results
previously published in \cite{Antonanzas-Torres.Canizares.ea2013}.

\lstset{language=R,numbers=none}
\begin{lstlisting} 
R> load('Stations.RData')
R> UTM <- SpatialPointsDataFrame(Stations[,c(2,3)], Stations[,-c(2,3)],
+                               proj4string=CRS('+proj=utm +zone=30 +ellps=WGS84'))


R> vgmCMSAF <- variogram(GHImed ~ GHIcmsaf, UTM)
R> fitvgmCMSAF <- fit.variogram(vgmCMSAF, vgm(model='Nug'))

R> gModel <- gstat(NULL, id='G0yKrig',
+                 formula= GHImed ~ GHIcmsaf,
+                 locations=UTM, model=fitvgmCMSAF)

R> names(GHI2005a) <- 'GHIcmsaf'
R> G0yKrig <- interpolate(GHI2005a, gModel, xyOnly=FALSE)
\end{lstlisting}

\subsection{Analysis of the results}
\label{sec-1-7}

\begin{figure}[H]
  \centering
  \includegraphics[width=0.9\textwidth]{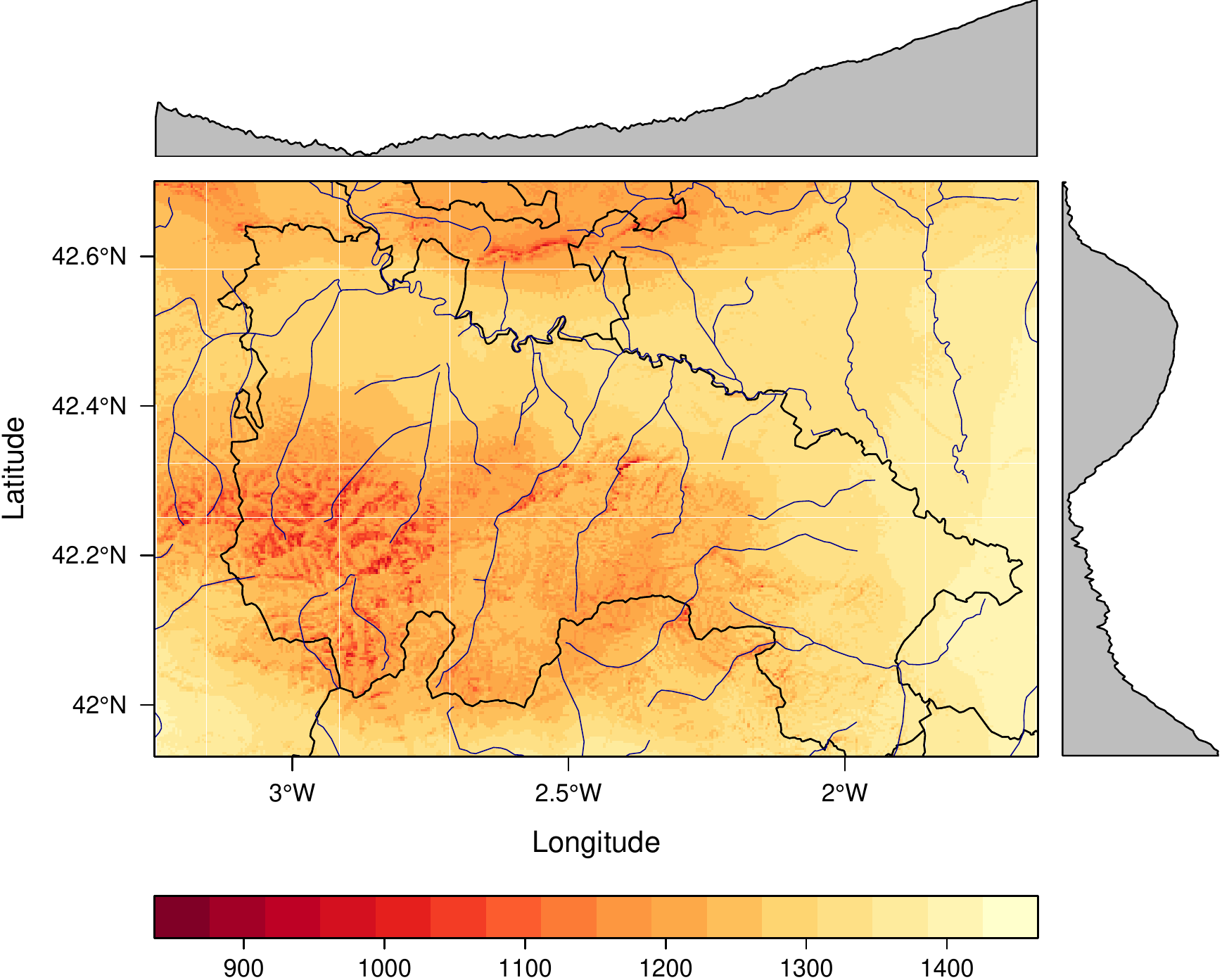}
  \caption{Annual GHI of 2005 downscaled without
    KED (0.03x0.03$^\circ$) in La Rioja}
  \label{fig:GHInoked}
\end{figure}

\begin{figure}[H]
  \centering
  \includegraphics[width=0.9\textwidth]{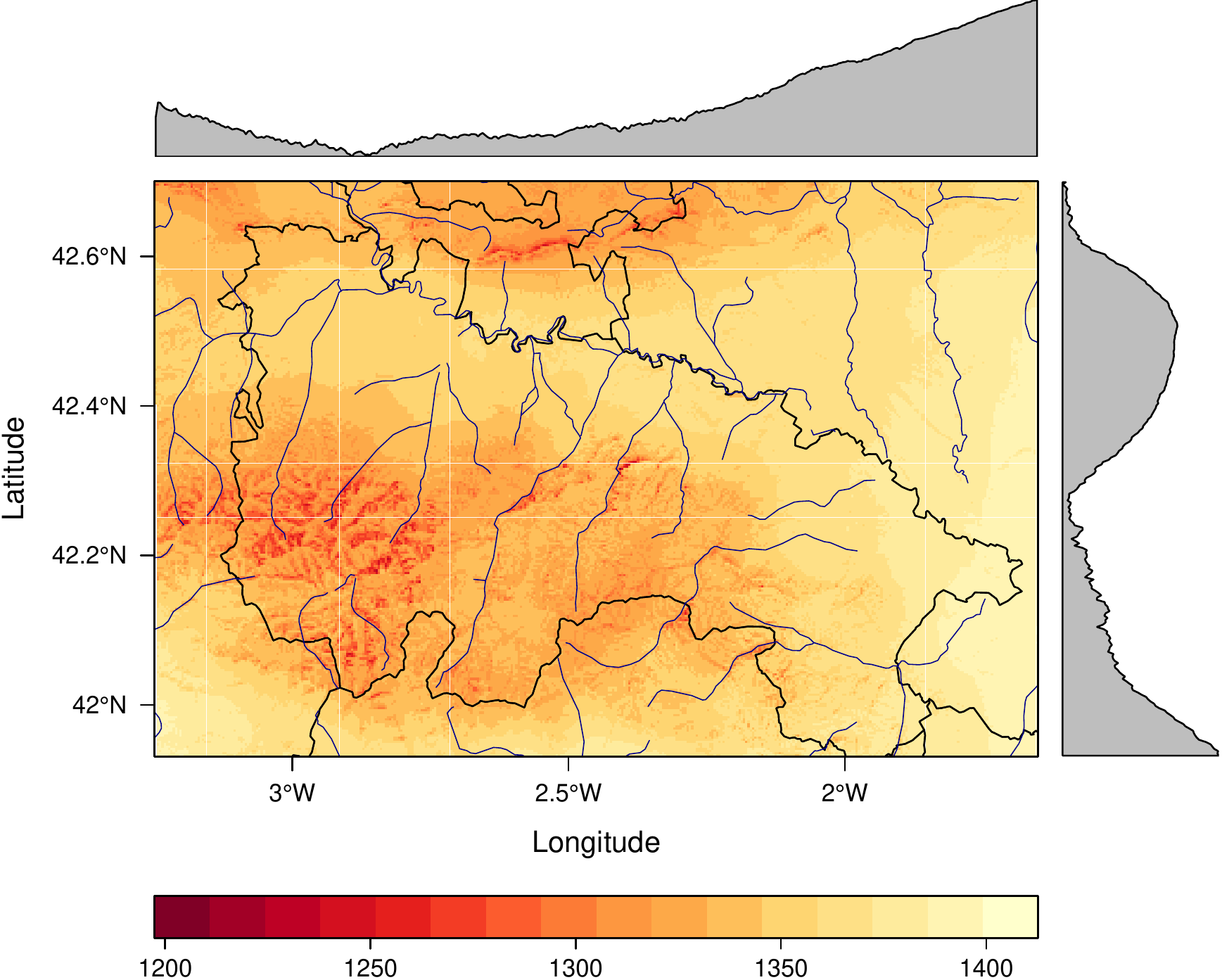}
  \caption{Annual GHI of 2005 downscaled with
    KED (0.03x0.03$^\circ$) in La Rioja}
  \label{fig:GHIked}
\end{figure}

Figure ~\ref{fig:cmsaf} shows the annual GHI as per CM~SAF with the
gross resolution analyzed (0.03x0.03$^\circ$) and
Figures~\ref{fig:GHInoked} and ~\ref{fig:GHIked} show the downscaled
maps (200x200 m) without and with the KED.

\subsubsection{Model performance}
\label{sec:error}

\begin{figure}[H]
  \centering
  \includegraphics[width=0.5\textwidth]{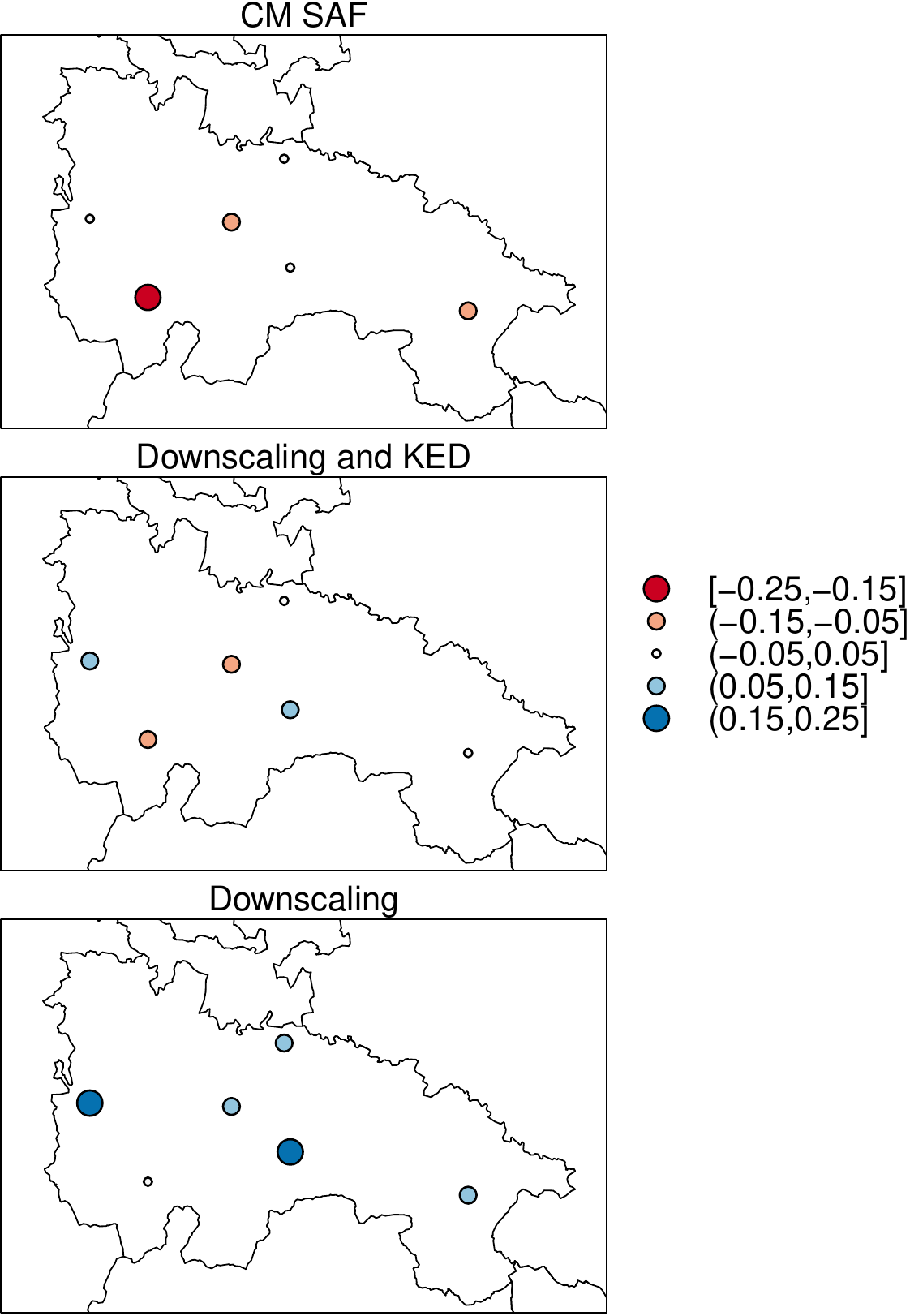}
  \caption{Annual relative differences evaluated with station measurements.}
  \label{fig:spplot}
\end{figure}

In order to evaluate the performance of the method proposed, relative
differences evaluated with station measurements are shown in
Figure \ref{fig:spplot}. As can be deduced from this Figure, relative
differences are smaller in \emph{downscaling with KED} than in CM SAF
or \emph{downscaling without KED}, at $\pm$ 15\%.  The mean absolute
error (MAE) and root mean square error (RMSE), described in Equations
\ref{eq:mae} and \ref{eq:rmse}, are used as indicators of model
performance.

\begin{equation}
\label{eq:mae}
MAE=\frac{\sum_{i=1}^n{\left |{x_{est}-x_{meas}}\right |}}{n} 
\end{equation}

\begin{equation}
\label{eq:rmse}
RMSE=\sqrt{\frac{\sum_{i=1}^n{(x_{est}-x_{meas})^2}}{n}}
\end{equation}
where \emph{n} is number of stations and $x_{est}$ and $x_{meas}$ the
annual estimated and measured irradiation, respectively.

Table ~\ref{tab:ers} shows the MAE and RMSE obtained with CM SAF and
with the methodology proposed before and after the KED. The KED leads
to a significant improvement in estimates: the MAE is down by 25.5\%
and the RMSE by 27.4\% compared to CM SAF.

\begin{table}[ht]
\begin{center}
\begin{tabular}{cccc}
  \hline
& CM SAF & without KED & with KED\\
\hline
MAE & 101.35 &  175.63& 75.54  \\ 
  RMSE & 118.65  & 196.53 & 86.18 \\ 
   \hline
\end{tabular}
\end{center}
\caption{Summary of errors obtained in $kWh/m^2$.}
\label{tab:ers}
\end{table}

The higher MAE recorded in station locations in CM SAF and
\emph{downscaling without KED} is also explained in the irradiation
maps shown in Figures ~\ref{fig:cmsaf} and ~\ref{fig:GHInoked}. The
$GHI_{annual}$ is lowered too far in certain regions of the area
studied with \emph{downscaling without KED} compared to
$GHI_{down,ked}$, which is also shown in Figure \ref{fig:spplot}.

\subsubsection{Zonal variability}
\label{sec:zonal-variation}

\begin{figure}[H]
  \centering
  \includegraphics[width=0.9\textwidth]{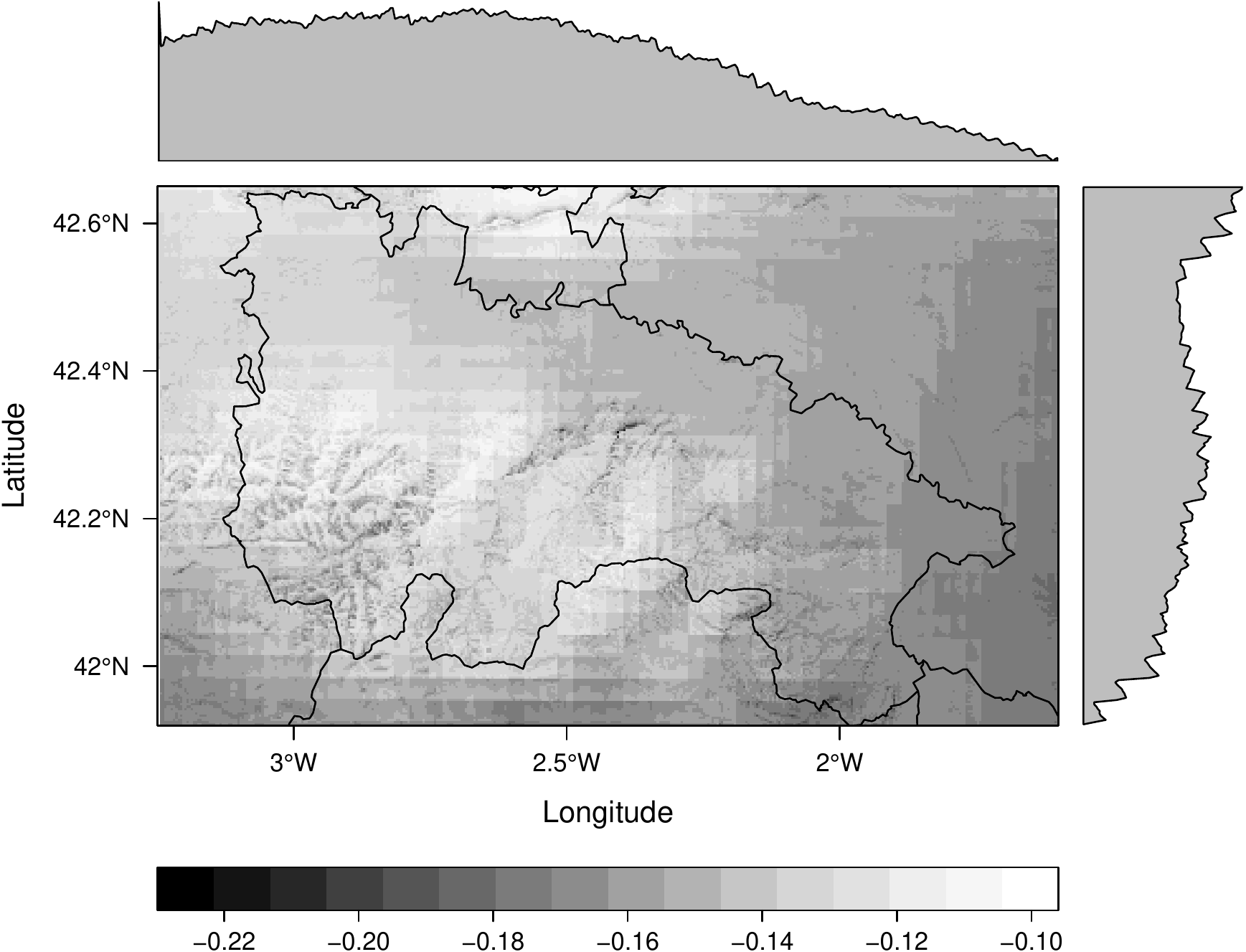}
  \caption{Relative difference of $GHI_{KED}$ and $GHI_{CMSAF, down}$  related to $GHI_{CMSAF, down}$}
  \label{fig:diffKEDcmsaf}
\end{figure}

\begin{figure}[H]
  \centering
  \includegraphics[width=0.9\textwidth]{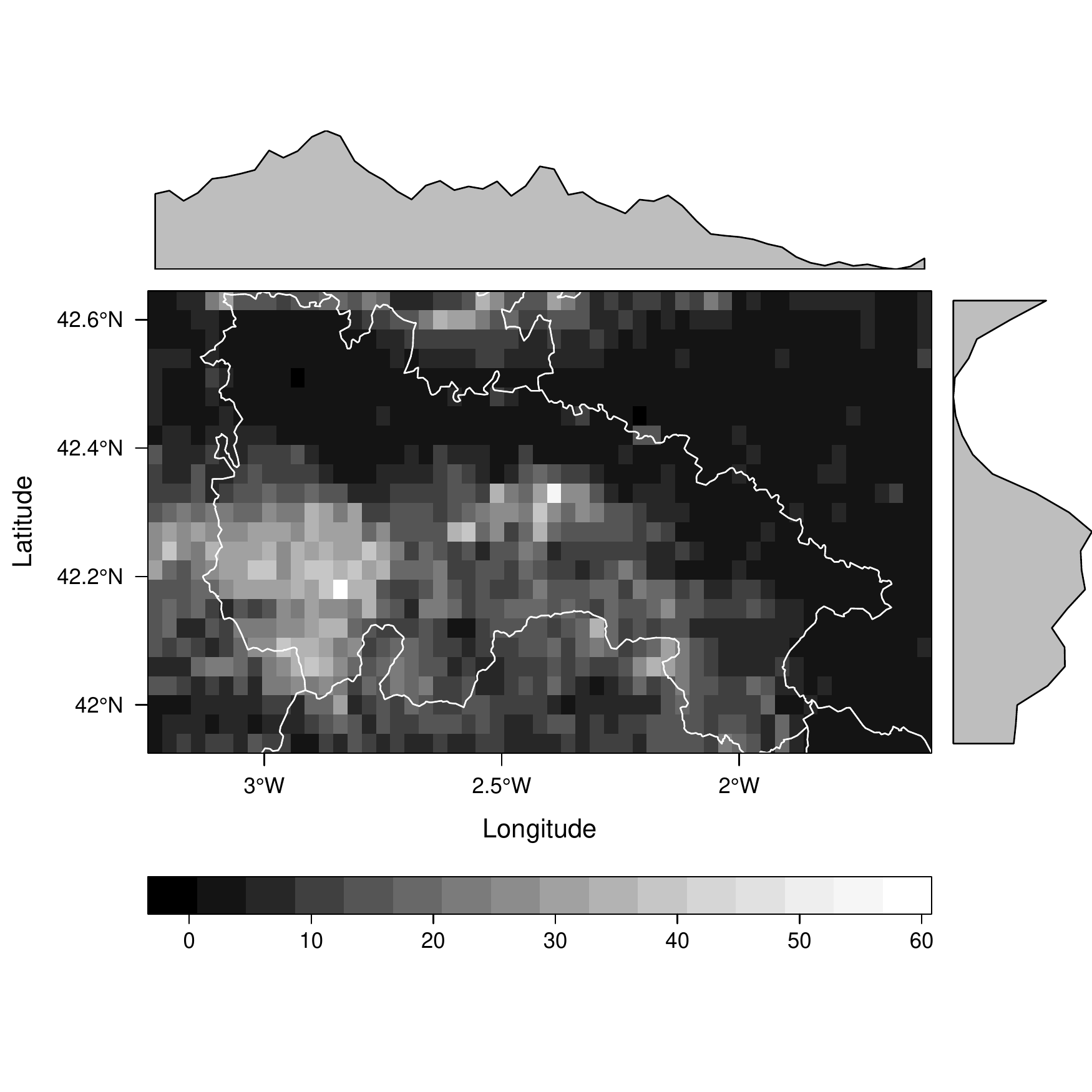}
  \caption{Difference of zonal standard deviations ($kWh/m^2$) between downscaling without KED and with KED.}
  \label{fig:zonal}
\end{figure}

\begin{figure}[H]
  \centering
  \includegraphics[width=0.7\textwidth]{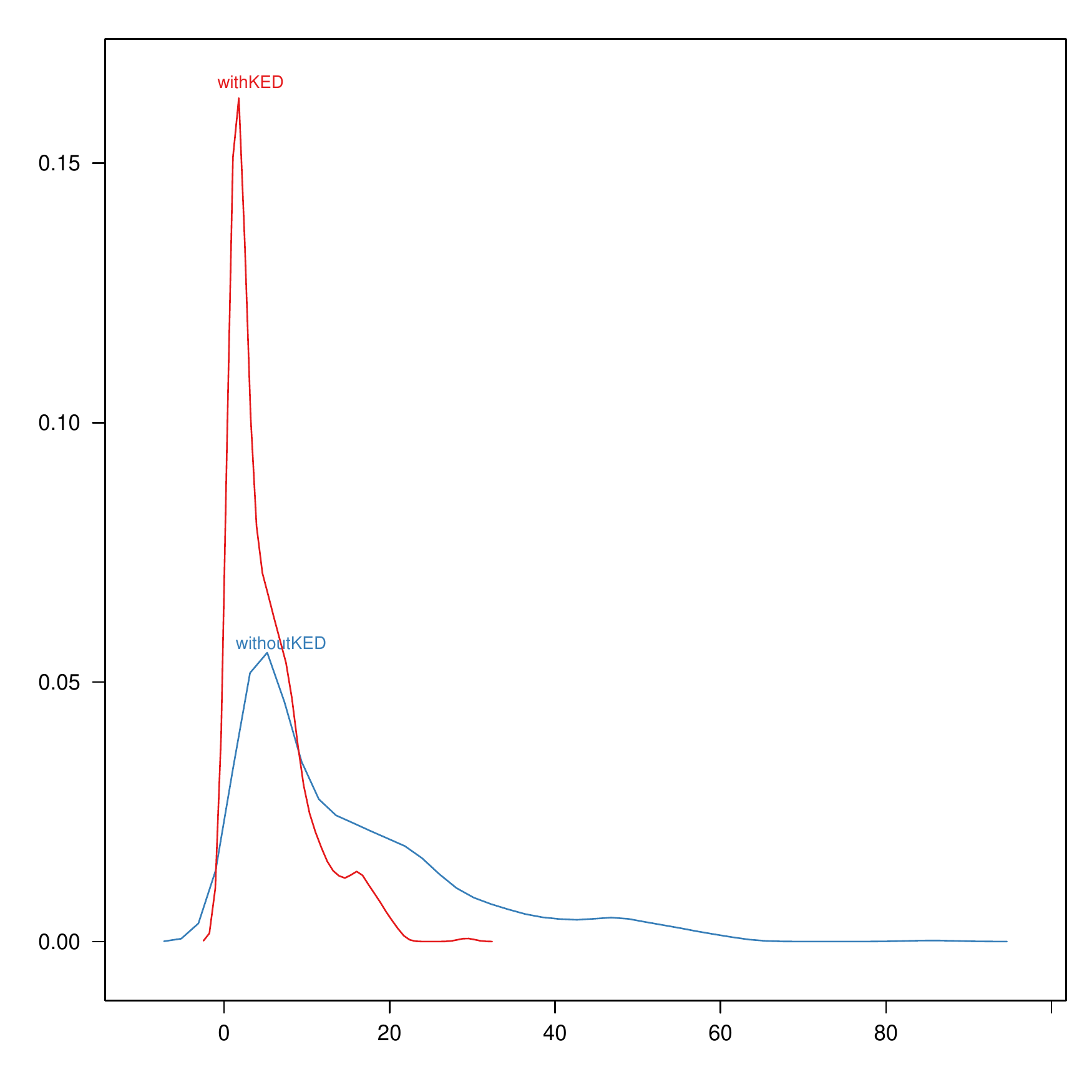}
  \caption{Density plot of zonal standard deviations between CM SAF and downscaling.}
  \label{fig:density}
\end{figure}

The intrapixel variability due to the downscaling procedure is
indicative of the importance of the topography as an attenuator of
solar irrradiation. As a result, this zonal variability is higher in
pixels with complex topographies and downscaling is more
useful. Figure \ref{fig:diffKEDcmsaf} shows the relative difference
between downscaling with KED and CM SAF. As might be deduced, CM SAF
over-estimates GHI in this region by between 11 and 22\%. Figures
~\ref{fig:zonal} and ~\ref{fig:density} display the standard
deviations of the downscaled maps within each cell of the original CM
SAF raster (0.03x0.03$^\circ$). The \texttt{zonal} function from the
\texttt{raster} library permits this calculation, explaining the
intrinsic variability of solar radiation within gross resolution
pixels. Consequently, in those pixels with higher standard deviations
there will be greater variability . Figure~\ref{fig:density} shows how
the KED method smooths the deviation within pixels and also the range
of solar irradiation in the region (Figures~\ref{fig:GHInoked}
and~\ref{fig:GHIked}).

\section{Concluding comments}

A methodology for downscaling solar irradiation is described and
presented using \texttt{R} software. This methodology is useful for
increasing the accuracy and spatial resolution of gross resolution
satellite-estimates of solar irradiation.

It has been proved that areas whose topography is complex show greater
differences with the original gross resolution data as a consequence
of horizon blocking and lower sky-view factors, so downscaling is
highly recommended in these areas.

\emph{Kriging with external drift} with the \texttt{gstat} package has
proved very useful in downscaling solar irradiation when on-ground
registers are available and an explanatory variable is provided.

This methodology is implemented as an example in the region of La
Rioja in northern Spain, and striking reductions of 25.5\% and 27.4\%
in MAE and RMSE are obtained compared to the original gross resolution
database. The high repeatability of this methodology and the reduction
in errors obtained might be also very useful in the downscaling of
meteorological variables other than solar irradiation.

\section*{Software information}
\label{sec:session}

The source code is available at
\url{https://github.com/EDMANSolar/downscaling}. The results discussed
in this paper were obtained in a \texttt{R} session with these
characteristics:

\begin{itemize}\raggedright
\item R version 2.15.2 (2012-10-26), \verb|x86_64-apple-darwin9.8.0|
  \item Locale:
    \verb|es_ES.UTF-8/es_ES.UTF-8/es_ES.UTF-8/C/es_ES.UTF-8/es_ES.UTF-8|
  \item Base packages: \texttt{base}, \texttt{datasets},
    \texttt{graphics}, \texttt{grDevices}, \texttt{grid},
    \texttt{methods}, \texttt{parallel}, \texttt{stats},\texttt{utils}
  \item Other packages: \texttt{foreign}~0.8-51,
    \texttt{gstat}~1.0-16, \texttt{hexbin}~1.26.0,
    \texttt{lattice}~0.20-15, \texttt{latticeExtra}~0.6-19,
    \texttt{maptools}~0.8-14, \texttt{raster}~2.1-16,
    \texttt{rasterVis}~0.20-01, \texttt{RColorBrewer}~1.0-5,
    \texttt{rgdal}~0.8-01, \texttt{solaR}~0.33, sp~1.0-8,
    \texttt{zoo}~1.7-9
  \item Loaded via a namespace (and not attached):
    \texttt{intervals}~0.14.0, \texttt{spacetime}~1.0-4,
    \texttt{tools}~2.15.2, \texttt{xts}~0.9-3
\end{itemize}

\section*{Acknowledgements}
We are indebted to the University of La Rioja (fellowship FPI2012) and
the Research Institute of La Rioja (IER) for funding parts of this
research.


\end{document}